\documentclass[aps,prd,preprint,preprintnumbers,nofootinbib,superscriptaddress]{revtex4}
\usepackage{ae}
\usepackage{bm} 
\usepackage{color}
\usepackage{amsmath}
\usepackage{amssymb}
\usepackage{amsfonts}
\usepackage{graphicx}
\usepackage{slashed}
\usepackage{setspace}
\usepackage{stackrel}
\usepackage{ulem}
\usepackage{color}

\newcommand{\Eqref}[1]{Eq.~\eqref{#1}}

\allowdisplaybreaks

\begin{document}

\setlength{\unitlength}{1mm}
\title{X-ray photon scattering at a focused high-intensity laser pulse}
\author{Felix Karbstein}\email{felix.karbstein@uni-jena.de}
\affiliation{Helmholtz-Institut Jena, Fr\"obelstieg 3, 07743 Jena, Germany}
\affiliation{Theoretisch-Physikalisches Institut, Abbe Center of Photonics, \\ Friedrich-Schiller-Universit\"at Jena, Max-Wien-Platz 1, 07743 Jena, Germany}
\author{Elena A. Mosman}\email{mosmanea@tpu.ru}
\affiliation{National Research Tomsk Polytechnic University, Lenin Ave. 30, 634050 Tomsk, Russia}
\affiliation{Physics Faculty, Tomsk State University, Lenin Ave. 36, 634050 Tomsk, Russia}

\date{\today}

\begin{abstract}
We study x-ray photon scattering in the head-on collision of an XFEL pulse and a focused high-intensity laser pulse, described as paraxial Laguerre-Gaussian beam of arbitrary mode composition.
For adequately chosen relative orientations of the polarization vectors of the colliding laser fields, this gives rise to a vacuum birefringence effect manifesting itself in polarization flipped signal photons.
Throughout this article the XFEL is assumed to be mildly focused to a waist larger than that of the high-intensity laser beam.
As previously demonstrated for the special case of a fundamental paraxial Gaussian beam, this scenario is generically accompanied by a scattering phenomenon of x-ray energy signal photons outside the forward cone of the XFEL beam, potentially assisting the detection of the effect in experiment.
Here, we study the fate of the x-ray scattering signal under exemplary deformations of the transverse focus profile of the high-intensity pump.
\end{abstract}

\maketitle

\section{Introduction}\label{sec:intro}

Quantum electrodynamics (QED) predicts effective nonlinear interactions of electromagnetic fields mediated by quantum fluctuations of electrons and positrons \cite{Euler:1935zz,Heisenberg:1935qt,Weisskopf,Schwinger:1951nm}. The coupling of the electromagnetic field to a virtual electron-positron loop is mediated by the elementary charge $e$.
In the low energy limit, i.e., for photon energies $\omega\ll m_e c^2$, where $m_e$ is the electron mass, these effective interactions are governed by the Heisenberg-Euler effective Lagrangian \cite{Heisenberg:1935qt}.
For field strengths much smaller than the critical electric and magnetic fields, $E_\text{cr}=m_e^2c^3/(e\hbar)\simeq 1.3 \times 10^{18}\,\frac{\rm V}{\rm m}$ and $B_\text{cr}=E_\text{cr}/c \simeq 4.4 \times 10^{9}\,{\rm T}$, respectively, the leading effective interaction amounts to a four-field interaction \cite{Euler:1935zz}. Higher-order interactions are parametrically suppressed with additional powers of $E/E_\text{cr}$ and $B/B_\text{cr}$, where $E$ ($B$) denotes the applied electric (magnetic) field.
These dimensionless ratios are much smaller than one for all present and near-future macroscopic electromagnetic fields available in the laboratory.
Correspondingly, a restriction to the leading interaction term is typically sufficient in studies of prospective signatures of QED vacuum nonlinearities with laboratory electromagnetic fields and optical to x-ray photon energies for probing.
As the fine-structure constant fulfills $\alpha=e^2/(4\pi\hbar \, c)\simeq1/137\ll1$, we can moreover limit our considerations to the one-loop level; cf., e.g., Ref.~\cite{Gies:2016yaa}.
For reviews, see Refs.~\cite{Dittrich:1985yb,Dittrich:2000zu,Marklund:2008gj,Heinzl:2008an,DiPiazza:2011tq,Dunne:2012vv,Battesti:2012hf,King:2015tba,Karbstein:2016hlj,Inada:2017lop}.

The collision of x-ray photons and a high-intensity laser field constitutes one of the most prospective scenarios towards the first experimental verification of QED vacuum nonlinearities in macroscopic electromagnetic fields in the laboratory.
The dominant signal in the x-ray domain arises from the quasi-elastic interaction of the incident x-ray photons with the oscillation period-averaged intensity profile of the high-intensity laser; cf. Ref.~\cite{Karbstein:2016lby} for a detailed discussion.
On the theoretical side, it is conveniently analyzed on the basis of the photon polarization tensor evaluated in the high-intensity laser field, mediating between an incident and outgoing photon line.
Upon identification of the incident photon line with the incident x-ray photons, we obtain a photon current, whose leading contribution is quadratic in the high-intensity laser field and linear in the x-ray field.
This current sources x-ray energy signal photons whose propagation and polarization properties are governed by the effective interaction of the colliding photon fields.

The original theoretical proposal to perform such an experiment \cite{Heinzl:2006xc} exclusively focused on the effect of polarization flipping induced by a birefringence phenomenon of the quantum vacuum subjected to a macroscopic electromagnetic field; cf. also Refs.~\cite{Dinu:2013gaa,Schlenvoigt:2016} for more detailed analyses of the effect.
Vacuum birefringence \cite{Toll:1952,Baier,BialynickaBirula:1970vy,Adler:1971wn} is already actively searched for in experiments using macroscopic magnetic fields to drive the effect and continuous wave laser for probing \cite{Cadene:2013bva,DellaValle:2015xxa,Fan:2017fnd}; see Ref.~\cite{Battesti:2018bgc} for a recent review.
As the birefringence signal is inversely proportional to the wavelength of the probe and directly proportional to the number of photons available for probing, employing an x-ray probe seems a very promising alternative route to verify the effect in experiment.

In the meantime, it has been demonstrated \cite{Karbstein:2015xra,Karbstein:2016lby,Karbstein:2018omb} that the scattering of signal photons outside the forward cone of the x-ray beam can be used to significantly increase the signal-to-background ratio. In principle, even the scattering phenomenon alone constitutes a signature of quantum vacuum nonlinearity.
The main idea is to make use of the different decay behavior of the photons constituting the driving x-ray beam and the signal photons as a function of the polar angle measured with respect to the beam axis of the x-ray beam.
While the former is determined by the focusing optics, the latter is controlled by the details of the scattering process, most specifically the transverse profile of the high-intensity beam in the interaction region.
The interaction region is the space-time volume where both pulses overlap, and ideally reach their peak field amplitudes.
Generically, the weaker the x-ray beam is focused relative to the high-intensity laser beam and thus the wider its waist, the more pronounced is the difference in the decay behavior \cite{Karbstein:2018omb}.
This can be intuitively understood by the fact that the weaker a beam is focused, the smaller its angular divergence. 
On the other hand, the smaller the scatterer, the wider the angular divergence of the scattering signal.
Performing such an experiment at the European XFEL \cite{XFEL} amounts to a central science goal of the HIBEF consortium \cite{HIBEF}, and is also targeted at SACLA \cite{Inada:2017lop} as well as SEL \cite{Shen:2018}.
Employing this scattering phenomenon might be essential for measuring the effect of vacuum birefringence at an XFEL equipped with a petawatt-class high-intensity laser and state-of-the-art high-purity x-ray polarimetry \cite{Marx:2011,Marx:2013xwa}.

So far, all available studies of the effect have modeled the high-intensity laser pump as fundamental paraxial Gaussian beam.
However, the details of the scattering process may depend sensitively on the precise spatio-temporal structure of the pump field, and in particular on its transverse intensity profile in the interaction region.
The present article is devoted to a first exploratory study of the fate of the scattering signal under deformations of the transverse focus profile of the pump.
To this end, we focus on the head-on collision of a weakly focused XFEL probe pulse with a tightly focused high-intensity laser pump pulse.
More specifically, throughout this article we assume the waist of the x-ray probe to be wider than that of the pump, thereby ensuring that the x-ray probe photons illuminate -- and thus probe -- the full transverse pump profile. 
While the former is modeled as linearly polarized plane wave supplemented with a finite pulse duration as in \cite{Karbstein:2015xra}, the latter is described as pulsed paraxial Laguerre-Gaussian (LG) beam of arbitrary mode composition \cite{Siegman,SalehTeich}, allowing us to consider generic rotationally symmetric beam profiles spanned by the LG basis.
The high-intensity laser field enters the calculation via the photon polarization tensor, which has been evaluated in the background of a pulsed LG beam in Ref.~\cite{Karbstein:2017jgh} based on a locally constant field approximation of the Heisenberg-Euler effective Lagrangian.

Our article is organized as follows:
In Sec.~\ref{sec:theory} we recall our theoretical treatment of the effect based on the approach adopted to the study of the vacuum birefringence signal induced by a fundamental paraxial Gaussian laser pulse in Ref.~\cite{Karbstein:2015xra}.
Our main focus is on the generalization of the results of Ref.~\cite{Karbstein:2015xra} to paraxial LG laser pulses of arbitrary mode composition, while -- at the same time -- keeping the resulting expressions as simple as possible.
In order to achieve this, we make use of several well-justified analytical approximations.
Thereafter, in Sec.~\ref{sec:results} we study the impact of differently shaped high-intensity laser pulse profiles on the scattering of the signal photons.
Finally, we end with Conclusions and an Outlook in Sec.~\ref{sec:concls}

\section{Theoretical Setting}\label{sec:theory}

In the Heaviside-Lorentz System with units $c=\hbar=1$, the leading effective interaction between slowly varying electromagnetic fields induced by QED vacuum fluctuations  is given by \cite{Euler:1935zz,Heisenberg:1935qt}
\begin{equation}
 {\cal L}_\text{HE}^{1\text{-loop}}\simeq\frac{m_e^4}{(8\pi)^2}\frac{1}{90}\Bigl(\frac{e}{m_e^2}\Bigr)^4\Bigl[4(F_{\mu\nu}F^{\mu\nu})^2+7(F_{\mu\nu} {}^\star\!F^{\mu\nu})^2\Bigr] ,
 \label{eq:Lag} 
\end{equation}
with field strength $F^{\mu\nu}=\partial^\mu A^\nu-\partial^\nu A^\mu$ and dual field strength tensor ${}^\star\!F^{\mu\nu}=\frac{1}{2}\epsilon^{\mu\nu\alpha\beta}F_{\alpha\beta}$, respectively; our metric convention is $g_{\mu \nu}=\mathrm{diag}(-1,+1,+1,+1)$.
Slowly varying fields vary on scales much larger than the reduced Compton wavelength (time) $\lambdabar_{\rm C}=\hbar/m_e\approx3.86 \cdot 10^{-13}\,{\rm m}$ ($\tau_{\rm
  C}=\lambdabar_{\rm C}/c\approx1.29 \cdot 10^{-21}\,{\rm s}$) of the electron.
  
The associated photon polarization tensor follows straightforwardly as \cite{Karbstein:2015cpa}
\begin{equation}
 \Pi^{\mu\nu}(-k',k)=\frac{\delta^2\int{\rm d}^4x\,{\cal L}^{1\text{-loop}}_\text{HE}}{\delta A_\mu(-k') \delta A_\nu(k)} \,.
 \label{eq:Pi}
\end{equation}
It is linear in $\alpha$, quadratic in the slowly varying electromagnetic field and mediates between incident $k^\mu=(k^0,\vec{k})$ and outgoing $k'^\mu=(k'^0,\vec{k}')$ photon momenta, fulfilling $\{k^0,k'^0,|\vec{k}|,|\vec{k}'|\}\ll m_e$.
Our conventions are such that both arguments of $\Pi^{\mu\nu}(.,.)$ are ingoing, resulting in the minus sign in the first argument of the polarization tensor in \Eqref{eq:Pi}.
Upon identification of this electromagnetic field with the field of a pulsed paraxial Laguerre-Gaussian laser beam of arbitrary mode composition, the photon polarization tensor encodes the vacuum-fluctuation-mediated impact of this background field on probe photon propagation.
In principle, higher order contributions in the background field (fine structure constant) could be systematically accounted for by extending the perturbative weak field (loop) expansion in \Eqref{eq:Lag}.
On the other hand, going beyond the restrictions of slowly varying fields and low energy photons, would eventually require an explicit calculation of the photon polarization tensor at arbitrary momentum transfer through the electron-positron loop in the specific inhomogeneous background field.
For explicit expressions of the photon polarization tensor in linearly (circularly) polarized paraxial Laguerre- and Hermite Gaussian beams obtained along the above lines, cf. Ref.~\cite{Karbstein:2017jgh} (\cite{Karbstein:2017uzq}).

Generic paraxial solutions of the wave equation in vacuum describing focused beams which feature a rotational symmetry around the beam axis can be represented as a superposition of Laguerre-Gaussian modes LG$_{l,p}$ labeled by two integer indices $l$ and $p$ \cite{Siegman,SalehTeich}.
They provide a convenient means to analytically model the electromagnetic fields of a focused high-intensity laser beam.
For a laser beam propagating in $\hat{\vec{e}}_{\rm z}$ direction, the directions of the associated electric and magnetic fields can be parameterized by a single angle $\phi$, and are given by $\hat{\vec{e}}_{E}=(\cos\phi,\sin\phi,0)$ and $\hat{\vec{e}}_{B}=\hat{\vec{e}}_{E}|_{\phi\to\phi+\frac{\pi}{2}}$.
Without loss of generality, we assume the beam to be focused at $\vec{x}=0$.
Linearly polarized beams fulfill $\phi=\text{const}.$ and are characterized by a single field profile $E(x)$, such that their electric and magnetic fields are given by $\vec{E}(x)=E(x)\hat{\vec{e}}_E$ and $\vec{B}(x)=E(x)\hat{\vec{e}}_B$, respectively.
In cylindrical coordinates ($r,\varphi,{\rm z}$), the latter is given by
\begin{equation}
 E(x)=\sum_{l,p}{\mathfrak E}_{l,p} \,{\rm e}^{-(\frac{{\rm z}-t}{\tau/2})^2}\biggl(\frac{\sqrt{2}r}{w({\rm z})}\biggr)^{|l|} L^{|l|}_p\Bigl(\bigl(\tfrac{\sqrt{2}r}{w({\rm z})}\bigr)^2\Bigr)\,\frac{w_0}{w({\rm z})}\,{\rm e}^{-(\frac{r}{w({\rm z})})^2}\,
 \cos\bigl(\Phi_{l,p}(x)+\varphi_{l,p}\bigr)\,,
 \label{eq:E}
\end{equation}
with
\begin{equation}
 \Phi_{l,p}(x)=\Omega({\rm z}-t)+\frac{\rm z}{{\rm z}_R} \Bigl(\frac{r}{w({\rm z})}\Bigr)^2-(|l|+2p+1)\arctan\Bigl(\frac{\rm z}{{\rm z}_R}\Bigr)-l\varphi\,.
 \label{eq:Phi}
\end{equation}
The radial focus profile is $E(r):=E(x)|_{{\rm z}=t=0}$.
Here, $\Omega$ is the oscillation frequency of the beam, $w_0$ its waist and ${\rm z}_R=\frac{\pi w_0^2}{\lambda}$ its Rayleigh range; $w({\rm z})=w_0\sqrt{1+(\frac{\rm z}{{\rm z}_R})^2}$. 
The sums over $l$ and $p$ run over all modes constituting the beam.
The peak field amplitude of a given mode ($l,p$) is ${\mathfrak E}_{l,p}$, and $\varphi_{l,p}$ denote mode specific phase factors.
To describe high-intensity laser pulses of finite energy, we have moreover supplemented the paraxial beam solution by a Gaussian temporal pulse profile of duration $\tau$, ensuring that it reaches its maximum in the beam focus at $t={\rm z}=0$.
This {\it ad hoc} modification gives rise to violations of Maxwell's equations in vacuum of ${\cal O}(\frac{1}{\tau\Omega})$, and hence should yield reliable results for $\tau\Omega\gg1$ \cite{Karbstein:2015cpa,Karbstein:2017jgh}.
The latter criterion is typically met for state-of-the-art high-intensity laser pulses.
On the other hand, the paraxial approximation neglects terms of ${\cal O}(\theta)$, where $\theta\simeq\frac{w_0}{{\rm z}_R}$ is the radial beam divergence in the far-field. 

Correspondingly, the photon polarization tensor in this field configuration constitutes the central input for the present study.
More specifically, throughout this article we only account for the quasi-elastic contribution to the photon polarization tensor characterized by $k'^0\approx k^0$, and neglect manifestly inelastic contributions, which depend on the oscillation frequency $\Omega$ of the high-intensity laser.
The latter are characterized by a finite energy transfer of the order of $2\Omega$, such that $k'^0\approx k^0\pm2\Omega$, corresponding to the absorption/release of two high-intensity laser photons by the x-ray probe.
These inelastic contributions are substantially suppressed as compared to the elastic one for the x-ray photon scattering scenario discussed here; cf. also Refs.~\cite{Karbstein:2015xra,Karbstein:2016lby}.
To keep this article self-contained, we reproduce the result \cite{Karbstein:2017jgh} for the photon polarization tensor in a linearly polarized pulsed LG beam here.
Taking into account the quasi-elastic contribution only, it can be represented as
\begin{equation}
 \Pi^{\rho\sigma}(-k',k) = \frac{\alpha}{\pi}\frac{1}{45}
   \Bigl[ 4\,(k'\hat F)^\rho  (k\hat F)^\sigma + 7\,(k'\,{}^\star\!\hat F)^\rho (k\,{}^\star\!\hat F)^\sigma \Bigr]\Pi(k-k') \,,
   \label{eq:Piexpl}
\end{equation}
with
\begin{align}
 \Pi(k-k') = &\sum_{l,p}\sum_{l',p'}\frac{e{\mathfrak E}_{l,p}}{m_e^2}\frac{e{\mathfrak E}_{l',p'}}{m_e^2}\,(2{\rm z}_R \, \pi w_0^2)\,
 \frac{\tau}{2}\sqrt{\frac{\pi}{2}}\,{\rm e}^{-\frac{1}{8}(\frac{\tau}{2})^2(k^0-k'^0)^2} \nonumber\\
 &\times\frac{1}{16}\sum_{j=0}^p\sum_{j'=0}^{p'}\frac{(-\sqrt{2})^{|l|+|l'|}}{ j!j'!}\binom{p+|l|}{p-j}\binom{p'+|l'|}{p'-j'} \nonumber\\
  &\times\sum_{\ell=\pm}{\rm e}^{ i  \ell(\varphi_{l,p}-\varphi_{l',p'})} \bigl[1+{\rm sign}\bigl(\ell(N-N')\bigr)\partial_{h_{\rm z}}\bigr]^{|N-N'|}\, \partial_c^{j+j'} \nonumber\\ 
 &\quad\times
 \bigl( i \partial_{h_{\rm x}}+{\rm sign}(l\ell)\partial_{h_{\rm y}}\bigr)^{|l|}\bigl( i \partial_{h_{\rm x}}-{\rm sign}(l'\ell)\partial_{h_{\rm y}}\bigr)^{|l'|} \nonumber\\  
 &\quad\times \frac{1}{c}F_{|N-N'|+|l|+|l'|}\bigl(h_{\rm z}-{\rm z}_R[(k-k')\hat\kappa],\tfrac{[w_0(\vec{k}_\perp-\vec{k}_\perp')+\vec{h}_\perp]^2}{8c}\bigr) \Big|_{c=1,\,\vec{h}=0}\,,
\end{align}
reflecting the fact that the leading contribution of the polarization tensor considered here is quadratic in the background field.
To keep this expression compact, we made use of the shorthand notations $\hat{\kappa}^\mu=(1,\hat{\vec{e}}_{\rm z})$, $\vec{k}_\perp=\vec{k}-(\vec{k}\cdot\hat{\vec{e}}_{\rm z})\hat{\vec{e}}_{\rm z}$, $N=|l|+2p$ 
and the following definition
\begin{align}
 F_\Lambda\bigl(a,b\bigr):&=\int_{-\infty}^\infty\frac{\rm dz}{{\rm z}_R}\,
 \Bigl(\frac{w_0}{w({\rm z})}\Bigr)^\Lambda \,{\rm e}^{- i a\frac{\rm z}{{\rm z}_R}-b\,(\frac{w({\rm z})}{w_0})^2} \nonumber\\
 &= \delta_{0,\Lambda}\sqrt{\frac{\pi}{b}}\,{\rm e}^{-\frac{1}{b}(\frac{a}{2})^2-b} + (1-\delta_{0,\Lambda}) \frac{\sqrt{\pi}}{\Gamma(\frac{\Lambda}{2})}\int_0^\infty\frac{{\rm d}s}{s}\,\frac{s^{\frac{\Lambda}{2}}}{\sqrt{s+b}}\,{\rm e}^{-\frac{1}{s+b}(\frac{a}{2})^2-(s+b)} \,. \label{eq:F}
\end{align}
The representation in the lower line is particularly convenient when aiming at a numeric evaluation of the function.
For real-valued arguments $a$ and $b$ it is manifestly real-valued and does not exhibit an oscillatory behavior.
The four-vectors spanning tensor structure in \Eqref{eq:Piexpl} can be expressed as \cite{Karbstein:2015cpa}
\begin{align}
 (k\hat F)^{\mu}&=(k\hat\kappa)\hat{e}^\mu_E-(k \hat{e}_E)\hat\kappa^\mu\,, \nonumber\\
 (k\,{}^\star\!\hat F)^{\mu}&=(k\hat\kappa)\hat{e}^\mu_B-(k \hat{e}_B)\hat\kappa^\mu\,,
\end{align}
where $\hat{e}_E^\mu=(0,\hat{\vec{e}}_E)$ and $\hat{e}_B^\mu=(0,\hat{\vec{e}}_B)$.
Note that as  $\lim_{\tau\to\infty}\frac{\tau}{2}\sqrt{\frac{\pi}{2}}\,{\rm e}^{-\frac{1}{8}(\frac{\tau}{2})^2(k^0-k'^0)^2} =  2\pi\,\delta(k^0-k'^0)$, in the limit of an infinitely long pulse duration $\tau$ we recover a strictly elastic contribution.
Furthermore, note that the electric peak field strength of a given mode can be straightforwardly related to the energy put into that mode as \cite{Karbstein:2017jgh} 
\begin{equation}
 {\mathfrak E}_{l,p}^2\simeq 8\sqrt{\frac{2}{\pi}}\frac{p!}{(p+|l|)!}\frac{W_{l,p}}{\pi w_0^2\tau}\,.
 \label{eq:Equad}
\end{equation}
The total laser pulse energy is given by
\begin{equation}
 W=\sum_{l,p}W_{l,p}\,.
 \label{eq:W}
\end{equation}
For convenience, we also make use of the definition
\begin{equation}
 E_0^2\simeq 8\sqrt{\frac{2}{\pi}}\frac{W}{\pi w_0^2\tau}\,,
\end{equation}
corresponding to the peak field strength squared, or equivalently peak intensity, of a fundamental Gaussian pulse of energy $W$.

As in Ref.~\cite{Karbstein:2015xra}, we assume the incident x-ray probe to be linearly polarized and in the interaction volume to be described by 
\begin{equation}
 A_\nu(x)=\frac{1}{2}\frac{\cal E}{\omega}\epsilon_{\nu}(\hat{k}){\rm e}^{ i \omega(\hat{k}x+t_0)-(\frac{\hat{k}x+t_0}{T/2})^2} .
 \label{eq:xray}
\end{equation}
Here, ${\cal E}$ is the peak electric field amplitude, $T$ the pulse duration and $\omega$ the oscillation frequency.
The pulse is propagating in direction $\hat{\vec{k}}$ (the associated four-vector is $\hat{k}^\mu=(1,\hat{\vec{k}})$) and its polarization vector is $\epsilon_{\nu}(\hat{k})=(0,\vec{\epsilon}\,(\hat{k}))$.
Finally, $t_0$ accounts for a finite temporal offset; i.e., the peak field is reached at $t=\hat{\vec{k}}\cdot\vec{x}+t_0$.
Equation~\eqref{eq:xray} fulfills Maxwell's equations in vacuum up to corrections of ${\cal O}(\frac{1}{T\omega})$ and thus constitutes a viable approximation for pulse durations fulfilling $T\omega\gg1$.
In this limit, the peak field amplitude can be related to the probe mean intensity $\langle I\rangle$ as ${\cal E}=\sqrt{2\langle I\rangle}$, 
The mean intensity can be expressed as $\langle I\rangle=J\omega$ in terms of the probe photon current density $J=\sqrt{\frac{8}{\pi}}\frac{N_{\rm in}}{\sigma T}$, which measures the number $N_{\rm in}$ of x-ray photons available for probing per transverse area $\sigma$ and time interval $T$.

Formally, the field~\eqref{eq:xray} only depends on the longitudinal coordinate and is infinitely extended in the transverse directions.
On the other hand, the studied signal of quantum vacuum nonlinearity is only induced in the overlap region of both the x-ray and the high-intensity laser fields and its amplitude scales quadratic with the high-intensity field strength.
Hence, to a quite good approximation this field can nevertheless be considered as mimicking the field of a weakly focused fundamental-mode Gaussian x-ray probe field in the vicinity of its beam focus: the important criterion is that the waist $w_{\rm probe}$ is much wider than the focal spot of the high-intensity laser field.

The effective interaction of the x-ray probe~\eqref{eq:xray}  with the strong electromagnetic field of the high-intensity laser pulse accounted for in the photon polarization tensor~\eqref{eq:Piexpl}  gives rise to a signal photon current
\begin{equation}
 j^\mu(x')=\int\frac{{\rm d}^4x}{(2\pi)^4}\,\Pi^{\mu\nu}(x',x)A_\nu(x)\,.
 \label{eq:j_pos}
\end{equation}
Taking into account the explicit form of the probe field given in \Eqref{eq:xray}, in momentum space the latter can be expressed as
\begin{equation}
 j^\mu(k')=\sqrt{\frac{J}{2\omega}}\,{\cal M}^{\mu\nu}(-k',k)\epsilon_{\nu}(\hat{k})\,,
 \label{eq:j_mom}
\end{equation}
where we defined \cite{Karbstein:2015xra}
\begin{equation}
 {\cal M}^{\mu\nu}(-k',k)=\sqrt{\pi}\,\frac{T}{2}\,{\rm e}^{-(\frac{2t_0}{T})^2}\int\frac{{\rm d}\tilde\omega}{2\pi}\,{\rm e}^{-\frac{1}{4}(\frac{T}{2})^2(\tilde\omega-\omega)^2+it_0\tilde\omega}\,\Pi^{\mu\nu}(-k',\tilde\omega\hat{k})\,.
 \label{eq:M}
\end{equation}
Subsequently we will also make use of the following definition,
\begin{equation}
 {\cal M}(k-k')=\sqrt{\pi}\,\frac{T}{2}\,{\rm e}^{-(\frac{2t_0}{T})^2}\int\frac{{\rm d}\tilde\omega}{2\pi}\,{\rm e}^{-\frac{1}{4}(\frac{T}{2})^2(\tilde\omega-\omega)^2+it_0\tilde\omega}\,\frac{\tilde\omega}{\omega}\Pi(\tilde\omega\hat{k}-k')\,.
 \label{eq:Ms}
\end{equation}
For completeness, note that these quantities fulfill $\lim_{T\to\infty}{\cal M}^{\mu\nu}(-k',k)={\rm e}^{ i t_0\omega}\,\Pi^{\mu\nu}(-k',k)$ and analogously $\lim_{T\to\infty}{\cal M}(k-k')={\rm e}^{ i t_0\omega}\,\Pi(k-k')$ with $k^\mu=\omega\hat{k}^\mu$.

The amplitude for the emission of a signal photon of four-momentum $k'^\mu=\omega'(1,\hat{\vec{k}}')$ and polarization vector $\epsilon_\mu^{(p)}(\hat{k}')$ is given by ${\cal S}_{(p)}(k')=\frac{\epsilon_\mu^{*(p)}(\hat{k}')}{\sqrt{2\omega'}}j^\mu(k')$, wherefrom the associated  differential signal photon number far outside the interaction region follows as ${\rm d}^3N^{(p)}=\frac{{\rm d}^3k'}{(2\pi)^3}\bigl|{\cal S}_{(p)}(k')\bigr|^2$ \cite{Karbstein:2014fva}.
In turn, we obtain
\begin{equation}
 {\rm d}^3N^{(p)}=\frac{{\rm d}^3k'}{(2\pi)^3}\biggl|\frac{\epsilon_\mu^{*(p)}(\hat{k}')}{\sqrt{2\omega'}} {\cal M}^{\mu\nu}(-k',k) \frac{\epsilon_{\nu}(\hat{k})}{\sqrt{2\omega}}\biggr|^2 J\,.
 \label{eq:dNp}
\end{equation}

As we primarily aim at studying the effect of the transverse high-intensity laser profile in the focus on the signal, in the present article we exclusively limit ourselves to counter-propagating beams, i.e., the special case of $\hat{\vec{k}}^\mu=(1,-\hat{\vec{e}}_{\rm z})$.
Moreover, we assume a linearly polarized x-ray probe and without loss of generality choose it to be polarized in $\rm x$ direction, i.e., $\epsilon_\nu(\hat{k})=(0,\hat{\vec{e}}_{\rm x})$.
On the other hand, the signal photons can in principle be emitted in arbitrary directions.
We parameterize their emission directions in spherical coordinates as $\hat{\vec{k}}'=(\cos\varphi'\sin\vartheta',\sin\varphi'\sin\vartheta',-\cos\vartheta')$ and their polarization vectors as ${\epsilon}^{(p)}_\mu(\hat{k})=(0,\sin\beta'\,\hat{\vec{k}}'|_{\vartheta=\frac{\pi}{2},\varphi'\to\varphi'+\frac{\pi}{2}}-\cos\beta'\,\hat{\vec{k}}'|_{\vartheta'\to\vartheta'+\frac{\pi}{2}})$ by means of the additional angle parameter $\beta'$.
For this choice, \Eqref{eq:dNp} can be represented as
\begin{align}
 {\rm d}^3N^{(p)}
 =\ &\frac{{\rm d}^3k'}{(2\pi)^3}\Bigl(\frac{1}{90}\frac{\alpha}{\pi}\Bigr)^2 \omega'\omega\,
(1+\cos\vartheta')^2 \bigl[11\cos(\varphi'-\beta')-3\cos(\varphi'-\beta'-2\phi)\bigr]^2 \nonumber\\
 &\times\bigl|{\cal M}(k-k')\bigr|^2 J\,.
 \label{eq:dNpexpl}
\end{align}
Note, that the entire polarization dependence of both the x-ray probe (controlled by $\phi$) and the signal photons ($\beta'$) is encoded in the last term in the first line of \Eqref{eq:dNpexpl}.

As the high-intensity laser pulse is characterized by frequencies much smaller than the x-ray frequency, it can only weakly impact the kinematics of x-ray photons.
Correspondingly, the signal photons are mainly emitted in the forward direction of the x-ray probe and decay rapidly with the polar angle $\vartheta'\ll1$ \cite{Karbstein:2015xra,Karbstein:2016lby}.

\subsection{ Generic rotationally symmetric high-intensity pumps}

The total number of signal photons ${\rm d}^3N^{\rm tot}(k')=\sum_{p}{\rm d}^3N^{(p)}(k')$ attainable in a polarization insensitive measurement is obtained by summing \Eqref{eq:dNpexpl} over two orthogonal signal photon polarizations, differing  by an angle of $\frac{\pi}{2}$ in the parameter $\beta'$.
On the other hand, the effect of vacuum birefringence manifests itself in polarization-flipped signal photons ${\rm d}^3N^{\perp}(k')$.
The latter are characterized by the condition $\vec{\epsilon}(\hat{k})\cdot\vec{\epsilon}^{\,(p)}(\hat{k}' )=0$, which is enforced by choosing $\beta'=-\arctan(\cos\vartheta'\cot\varphi')$.
For $\vartheta'\ll1$, this simplifies to $\beta'\simeq(\varphi'\ {\rm mod}\ \pi)-\frac{\pi}{2}$.
In turn,  from \Eqref{eq:dNpexpl} we obtain 
\begin{align}
\left\{\begin{array}{c}
 {\rm d}^3N^{\rm tot} \\
 {\rm d}^3N^{\perp}
\end{array}\right\}
 \simeq\ \frac{{\rm d}^3k'}{(2\pi)^3}\Bigl(\frac{1}{45}\frac{\alpha}{\pi}\Bigr)^2 \omega'\omega
\left\{\begin{array}{c}
130-66\cos(2\phi) \\
9\sin^2(2\phi)
\end{array}\right\}\bigl|{\cal M}(k-k')\bigr|^2 J\,.
 \label{eq:dNtotperp}
\end{align}
Obviously, the induced signal photon numbers ${\rm d}^3N^{\rm tot}$ and ${\rm d}^3N^{\perp}$ become maximum for different choices of the angle $\phi$.
While ${\rm d}^3N^{\rm tot}$ is maximized for an angle of $\phi=(\frac{\pi}{2}\ {\rm mod}\ \pi)$ between the polarization vectors of the counter-propagating electromagnetic beams, the optimal choice for ${\rm d}^3N^\perp$ is $\phi=(\frac{\pi}{4}\ {\rm mod}\ \frac{\pi}{2})$. 
Keeping all other parameters fixed, the corresponding maximum numbers fulfill $N^{\rm tot}_{\rm max}/N^\perp_{\rm max}=196/9\approx21.8$.

The integration over $\tilde\omega$ can be performed explicitly, such that the expression in the modulus in \Eqref{eq:dNtotperp} can be written as
\begin{align}
 {\cal M}(k-k') =\ &\pi\alpha\,\frac{2{\rm z}_R}{m_e^3\omega}\frac{\frac{T}{2}}{\sqrt{(\frac{T}{2})^2+\frac{1}{2}(\frac{\tau}{2})^2}}\,\Bigl(\omega+\frac{\tfrac{1}{2}(\tfrac{\tau}{2})^2\delta\omega+2( i t_0+2{\rm z}_R\partial_{h_{\rm z}})}{(\frac{T}{2})^2+\frac{1}{2}(\frac{\tau}{2})^2}\Bigr) \nonumber\\
 &\times{\rm e}^{-\frac{\frac{1}{2}(\frac{\tau}{2})^2}{(\frac{T}{2})^2+\frac{1}{2}(\frac{\tau}{2})^2}[(\frac{T}{2})^2(\frac{\delta\omega}{2})^2- i t_0\delta\omega]}\,
 {\rm e}^{\frac{(2{\rm z}_R)^2-t_0^2}{(\frac{T}{2})^2+\frac{1}{2}(\frac{\tau}{2})^2}-(\frac{2t_0}{T})^2+ i t_0\omega}
 \nonumber\\
 &\times\sum_{l,p}\sum_{l',p'} \,\sqrt{\frac{W_{l,p}}{m_e}\frac{W_{l',p'}}{m_e}}\sum_{j=0}^p\sum_{j'=0}^{p'}\frac{(-\sqrt{2})^{|l|+|l'|}}{ j!j'!} \nonumber\\
 &\times \,\frac{\sqrt{(p+|l|)!p!(p'+|l'|)!p'!}}{(p-j)!(|l|+j)!(p'-j')!(|l'|+j')!}\, \partial_c^{j+j'}  \nonumber\\
  &\times\sum_{\ell=\pm}{\rm e}^{ i  \ell(\varphi_{l,p}-\varphi_{l',p'})} \bigl[1+{\rm sign}\bigl(\ell(N-N')\bigr)\partial_{h_{\rm z}}\bigr]^{|N-N'|} \nonumber\\ 
 &\quad\times
 \bigl( i \partial_{h_{\rm x}}+{\rm sign}(l\ell)\partial_{h_{\rm y}}\bigr)^{|l|}\bigl( i \partial_{h_{\rm x}}-{\rm sign}(l'\ell)\partial_{h_{\rm y}}\bigr)^{|l'|} \nonumber\\  
 &\quad\times \frac{1}{c}F_{|N-N'|+|l|+|l'|}\bigl(\bar{a},\bar{b}\bigr) \bigg|_{c=1,\,\vec{h}=0},
 \label{eq:intdomegatilde}
\end{align}
with
\begin{align}
 \bar{a}&=h_{\rm z}+2{\rm z}_R\Bigl[\frac{1+\hat{k}_{\rm z}'}{2}(\omega+\delta\omega)-\frac{(\frac{T}{2})^2}{(\frac{T}{2})^2+\frac{1}{2}(\frac{\tau}{2})^2}\delta\omega+\frac{2 i t_0}{(\frac{T}{2})^2+\frac{1}{2}(\frac{\tau}{2})^2}\Bigr]\,, \nonumber\\
 \bar{b}&=\frac{[\vec{h}_\perp-w_0(\omega+\delta\omega)\hat{\vec{k}}_\perp']^2}{8c}+\frac{(2{\rm z}_R)^2}{(\frac{T}{2})^2+\frac{1}{2}(\frac{\tau}{2})^2}\,, \label{eq:bara+barb}
\end{align}
and $\omega'=\omega+\delta\omega$.
The Gaussian term $\sim{\rm e}^{-\#\delta\omega^2}$ in the second line of \Eqref{eq:intdomegatilde} ensures that the signal photon energy is peaked at $\omega'=\omega$ and falls off rapidly towards smaller and larger energies.
The associated $1/{\rm e}$ half width is $\delta\omega=2\sqrt{(\frac{T}{2})^2+\frac{1}{2}(\frac{\tau}{2})^2}/\sqrt{\frac{1}{2}(\frac{\tau}{2})^2(\frac{T}{2})^2}$.
To arrive at the expression in \Eqref{eq:intdomegatilde} we made use of the representation of $F_\Lambda(.,.)$ in the first line of \Eqref{eq:F}: performing the integration over $\tilde\omega$, the result can again be expressed in terms of $F_\Lambda(.,.)$ with shifted arguments.
 Equations~\eqref{eq:dNtotperp}-\eqref{eq:bara+barb} constitute our most general results.

\subsection{Analytical insights for high-intensity pumps with $l=0$}\label{subsec:approx}

In the remainder of this article we exclusively focus on the special case of pump fields spanned by LG$_{0,p}$ modes only, i.e., consider the sector characterized by $l=l'=0$.
This amounts to the sector of light without orbital angular momentum  \cite{Allen:1992zz}.
Besides we consider only optimal collisions with vanishing temporal offset, i.e., set $t_0=0$.
In this case, \Eqref{eq:intdomegatilde} simplifies significantly and features only parameter differentiations for $c$ and $h_{\rm z}$, allowing us to set $\vec{h}_\perp=0$ from the outset.

 Additional simplifications are possible in the parameter regime where $\bar a$ is small.
From the representation of $F_\Lambda(.,.)$ given in the second line of \Eqref{eq:F} we infer that the precise criterion is $(\frac{\bar a}{2})^2/{\bar b}\ll1$.
Taking the above $1/{\rm e}$ half width as a measure of the typical spread of $\delta\omega$, this can be translated to the physical conditions
\begin{equation}
 \{\vartheta',\tau\omega\vartheta'^2,T/\tau\}\ll1 \quad \text{and} \quad \frac{\tau}{2{\rm z}_R}\lesssim{\cal O}(1) \,. \label{eq:approxconds}
\end{equation} 
For completeness, note that the conditions~\eqref{eq:approxconds} ensure $(\frac{\bar a}{2})^2/{\bar b}\ll1$ for all polar angle $0\leq\vartheta'\ll1$. Requiring this criterion to be met only for angles $\vartheta'_0\leq\vartheta'\ll1$ beyond a certain minimum value  $\vartheta'_0$, less restrictive conditions can be derived along the same lines.
Accounting only for the leading order terms, in this parameter regime the quantities defined in \Eqref{eq:bara+barb} simplify significantly and read
\begin{align}
 \bar{a}\simeq h_{\rm z}\,, \quad \bar{b}\simeq\frac{(w_0\omega\vartheta')^2}{8c}+\frac{(2{\rm z}_R)^2}{(\frac{T}{2})^2+\frac{1}{2}(\frac{\tau}{2})^2}\,. \label{eq:barabarbsimp}
\end{align}
Moreover,  the terms in the round brackets in the first line of \Eqref{eq:intdomegatilde} become $\simeq\omega$.
As will be shown in Sec.~\ref{sec:results} below,  the above assumptions are fully compatible with typical XFEL and high-intensity laser parameters.

Under the above assumptions, \Eqref{eq:intdomegatilde} can be expressed as
\begin{align}
 {\cal M}(k-k')\simeq\ &2\pi\alpha\,\frac{2{\rm z}_R}{m_e^3}\frac{\frac{T}{2}}{\sqrt{(\frac{T}{2})^2+\frac{1}{2}(\frac{\tau}{2})^2}}\,
 {\rm e}^{-\frac{\frac{1}{2}(\frac{\tau}{2})^2}{(\frac{T}{2})^2+\frac{1}{2}(\frac{\tau}{2})^2}(\frac{T}{2})^2(\frac{\delta\omega}{2})^2}\,
 {\rm e}^{\frac{(2{\rm z}_R)^2}{(\frac{T}{2})^2+\frac{1}{2}(\frac{\tau}{2})^2}}\, \sum_{p,p'} {\cal I}_{p,p'}\,,
 \label{eq:intdomegatilde0}
\end{align}
where we introduced the shorthand notation
\begin{align}
 {\cal I}_{p,p'}=\ &\sqrt{\frac{W_{0,p}}{m_e}\frac{W_{0,p'}}{m_e}}\,\cos(\varphi_{0,p}-\varphi_{0,p'})  \nonumber\\
  &\times\sum_{j=0}^p\sum_{j'=0}^{p'} \binom{p}{j}\binom{p'}{j'}\frac{\partial_c^{j+j'}}{ j!j'!} \frac{1}{c}F_{2|p-p'|}\Bigl(0,\tfrac{(w_0\omega\vartheta')^2}{8c}+\tfrac{(2{\rm z}_R)^2}{(\frac{T}{2})^2+\frac{1}{2}(\frac{\tau}{2})^2}\Bigr) \bigg|_{c=1}.
  \label{eq:Ipp'}
\end{align}
Note, that \Eqref{eq:Ipp'} depends on the polar angle $\vartheta'$ only via the combination $w_0\omega\vartheta'$.
In the considered limit, the function $F_{\Lambda}(.,.)$ introduced in \Eqref{eq:F} has the following representation in terms of standard functions and the error function ${\rm erf}(.)$,
\begin{align}
 F_{2n}\bigl(0,b\bigr)= \delta_{0,n}\sqrt{\frac{\pi}{b}}\,{\rm e}^{-b} + (1-\delta_{0,n}) \frac{\pi}{(n-1)!}\,(-\partial_u)^{n-1}\frac{1}{\sqrt{u}}\,{\rm e}^{b(u-1)}\bigl[1-{\rm erf}(\sqrt{bu})\bigr]\bigg|_{u=1} \,, \label{eq:F0}
\end{align}
where $n=|p-p'|\in\mathbb{N}_0$; cf. formula 3.362.2 of \cite{Gradshteyn}.

In a next step we aim at integrating the differential numbers~\eqref{eq:dNtotperp} over the signal photon energy $\omega'$ and the azimuthal angle $\varphi'$. 
Within the accuracy of the above approximations, we have $\omega'{\rm d}^3k'\simeq \omega^3\vartheta'{\rm d}\vartheta'{\rm d}\omega'{\rm d}\varphi'$, and the limits of the Gaussian integration over the signal photon energy $\omega'$, or equivalently $\delta\omega$, can be formally extended to $\pm\infty$, allowing for a straightforward analytical integration.
Besides, the rotational symmetry of the considered head-on collision scenario around the common beam axis of the driving fields renders the integration over $\varphi'$ trivial.
The result for the differential numbers of signal photons is
\begin{align}
\left\{\begin{array}{c}
 {\rm d}N^{\rm tot} \\
 {\rm d}N^{\perp}
\end{array}\right\}
 \simeq\ &\vartheta'{\rm d}\vartheta'\Bigl(\alpha\frac{\omega}{m_e}\Bigr)^4\frac{T}{\tau}\Bigl(\frac{1}{45\pi}\frac{2{\rm z}_R}{m_e}\Bigr)^2
\left\{\begin{array}{c}
130-66\cos(2\phi) \nonumber\\
9\sin^2(2\phi)
\end{array}\right\} \\
&\times\frac{2\sqrt{\pi}}{\sqrt{(\frac{T}{2})^2+\frac{1}{2}(\frac{\tau}{2})^2}}\,
 \,
 {\rm e}^{\frac{2(2{\rm z}_R)^2}{(\frac{T}{2})^2+\frac{1}{2}(\frac{\tau}{2})^2}}\, \Bigl|\sum_{p,p'} {\cal I}_{p,p'}\Bigr|^2 J\,.
 \label{eq:dNtotperp_dcostheta}
\end{align}
Aiming at the explicit evaluation of this expression, it is convenient to decompose the sums over $p$ and $p'$ in \Eqref{eq:dNtotperp_dcostheta} into diagonal and off-diagonal parts as

\begin{equation}
 \sum_{p=0}^{\cal N}\sum_{p'=0}^{\cal N} {\cal I}_{p,p'} = \sum_{p=0}^{\cal N} {\cal I}_{p,p} + 2\sum_{p=0}^{\cal N}\sum_{p'=0}^{p-1} {\cal I}_{p,p'}\,.
\end{equation}
On the other hand, the far-field angular decay of a fundamental-mode Gaussian probe of waist $w_{\rm probe}$ encompassing $N_{\rm in}$ photons is given by
\begin{equation}
 \frac{{\rm d}N_{\rm in}}{{\rm d}\vartheta'\vartheta'}\simeq N_{\rm in}(w_{\rm probe}\omega)^2\,{\rm e}^{-\frac{1}{2}(w_{\rm probe}\omega\vartheta')^2}\,. \label{eq:dNin}
\end{equation}

\section{Exemplary results}\label{sec:results}

Subsequently, we aim at studying the effect of different transverse profiles of the high-intensity laser on the scattering signal.
To this end, we briefly focus on Gaussian laser fields prepared in a pure LG mode with $l=0$, and on the class of flattened Gaussian beams introduced in Ref.~\cite{Gori:1994}, which closely resemble super-Gaussian laser fields \cite{Palma:1994}.

In our explicit examples, we stick to the experimentally realistic parameters adopted in Ref.~\cite{Karbstein:2015xra} for the study of vacuum birefringence in the collision of an XFEL probe pulse with a fundamental-mode (${\rm LG}_{0,0}$) Gaussian high-intensity pump.
To be specific, we assume the XFEL to deliver pulses of duration $T=10\,{\rm fs}\simeq15.20\,{\rm eV}^{-1}$, comprising $N_{\rm in}=10^{12}$ photons at an energy of $\omega=12914\,{\rm eV}$.
The polarization purity of x-ray photons of this energy can be measured to the level of ${\cal P}=5.7\times10^{-10}$ \cite{Marx:2011,Marx:2013xwa}. 
For the parameters of the counter-propagating high-intensity laser field we assume a pulse energy of $W=30\,{\rm J}\simeq1.87\times10^{20}\,{\rm eV}$, a pulse duration of $\tau=30\,{\rm fs}\simeq45.60\,{\rm eV}^{-1}$ and a wavelength of $\lambda=800\,{\rm nm}\simeq4.06\,{\rm eV}^{-1}$, representing a typical state-of-the-art high-intensity laser system of the petawatt class.
As a realistic estimate for the beam waist of the fundamental mode we choose $w_0=1000\,{\rm nm}\simeq5.07\,{\rm eV}^{-1}$. The corresponding Rayleigh range is ${\rm z}_R\simeq19.89\,{\rm eV}^{-1}$.
The radius of the probe $w_\text{probe}$ determining the transverse area $\sigma=\pi w_\text{probe}^2$ is assumed to be given by $w_\text{probe}=3w_0$.

For these parameters we have $\tau\omega\simeq 5.89\times10^{5}$, such that the criterion $\{\vartheta',\tau\omega\vartheta'^2\}\ll1$ adopted in the derivation of the compact approximate expression for the differential signal photon number~\eqref{eq:dNtotperp_dcostheta} translates into the constraint $\vartheta'\ll0.001\,{\rm rad}$.
Because of $T/\tau=\frac{1}{3}$ and $\frac{\tau}{2{\rm z}_R}\simeq1.15$, the additional criteria are also well met, justifying the use of the approximate expressions in the exemplary studies presented below. 
As a further indication of the applicability of the approximate expressions in the considered parameter regime, see the good agreement of the results of a direct numerical evaluation of Eqs.~\eqref{eq:dNtotperp}-\eqref{eq:bara+barb} and the analytical approximation~\eqref{eq:dNtotperp_dcostheta} highlighted in Fig.~\ref{fig:LG} for the case of an ${\rm LG}_{1,0}$ pump.
Apart from this, all the results presented in the remainder of this article are based on the approximate expression~\eqref{eq:dNtotperp_dcostheta} for the differential number of attainable signal photons detailed in Sec.~\ref{subsec:approx}.

\subsection{Pure Laguerre-Gaussian modes}\label{subsec:LG}

For the special case of the high-intensity laser field corresponding to a pure LG{$_{0,p}$} mode, characterized by $l=0$ and a fixed value of $p\in\mathbb{N}_0$, we have
\begin{align} 
 \sum_{p,p'}{\cal I}_{p,p'}\to{\cal I}_{p,p}=\ &\frac{W_{0,p}}{m_e}\sum_{j=0}^p\sum_{j'=0}^{p} \binom{p}{j}\binom{p}{j'}\frac{\partial_c^{j+j'}}{ j!j'!}  \nonumber\\
  &\times \frac{\sqrt{\pi}}{c}\,\Bigl(\frac{(w_0\omega\vartheta')^2}{8c}+\frac{(2{\rm z}_R)^2}{(\frac{T}{2})^2+\frac{1}{2}(\frac{\tau}{2})^2}\Bigr)^{-\frac{1}{2}}\,{\rm e}^{-\frac{(w_0\omega\vartheta')^2}{8c}-\frac{(2{\rm z}_R)^2}{(\frac{T}{2})^2+\frac{1}{2}(\frac{\tau}{2})^2}} \bigg|_{c=1},
  \label{eq:Ipp}
\end{align}
with $W_{0,p}=W$.
Upon insertion of this expression into \Eqref{eq:dNtotperp_dcostheta}, it is straightforward to determine the differential number of x-ray signal photons arising from the effective interaction of the x-ray probe with the counter-propagating LG high-intensity laser pulse. 
\begin{figure}[h]
 \centering
\includegraphics[width=0.67\columnwidth]{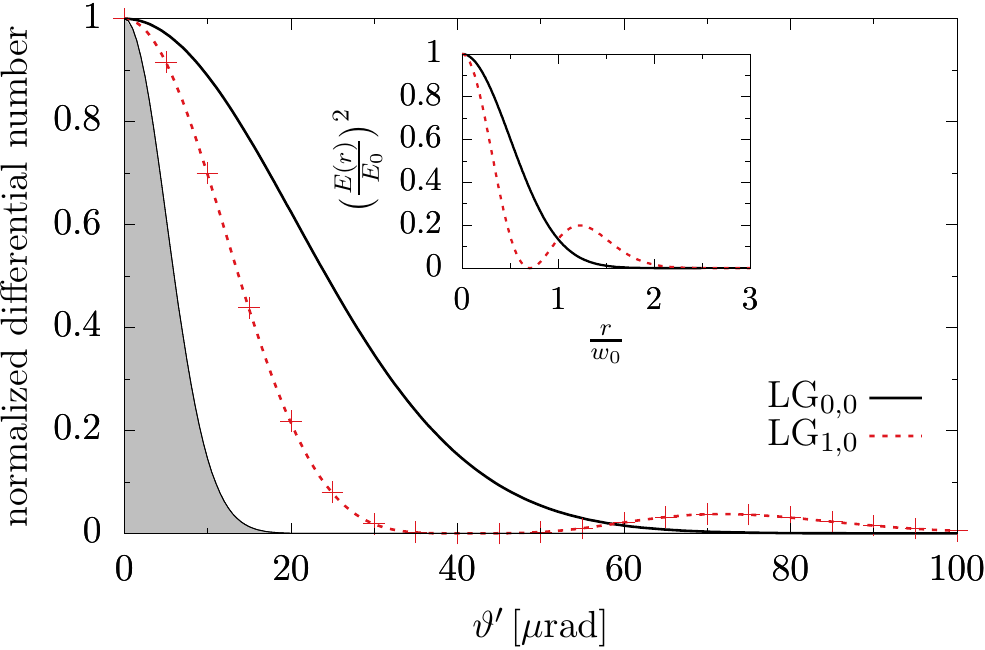}
\caption{Far-field angular decay of the signal~\eqref{eq:dNtotperp_dcostheta} as a function of the polar angle $\vartheta'$ measured from the forward beam axis of the probe for different transverse pump profiles.
Here, we consider the pump to be prepared in a pure LG mode; see the inset for the radial intensity profiles in the focus.
For comparison, in gray we also highlight the angular decay~\eqref{eq:dNin} of a fundamental Gaussian probe of waist $w_{\rm probe}=3w_0$; $w_0$ is the waist of the fundamental pump mode ${\rm LG}_{0,0}$. The plot is for $\omega=12914\,{\rm eV}$, $\lambda=800\,{\rm nm}$, $w_0=1000\,{\rm nm}$, $T=10\,{\rm fs}$ and $\tau=30\,{\rm fs}$.
This plot also highlights the quality of our analytical approximation~\eqref{eq:dNtotperp_dcostheta}: the data points marked by crosses follow from a direct numerical evaluation of Eqs.~\eqref{eq:dNtotperp}-\eqref{eq:bara+barb} for the ${\rm LG}_{1,0}$ scenario. They are in good agreement with the red dashed line obtained from \Eqref{eq:dNtotperp_dcostheta}.
}
\label{fig:LG}
\end{figure}
Figure~\ref{fig:LG} depicts the far-field angular decay of the signal photons~\eqref{eq:dNtotperp_dcostheta} as a function of the polar angle $\vartheta'$ for both the fundamental Gaussian mode ${\rm LG}_{0,0}$ studied already in detail in Ref.~\cite{Karbstein:2015xra} and the ${\rm LG}_{1,0}$ mode. As our focus is on normalized quantities, -- apart from the special case of $\sin(2\phi)=0$ -- the results for $N^{\rm tot}$ and $N^\perp$ are described by the same curves.

When giving explicit numeric results for the signal photon number $N^{\rm tot}$ ($N^\perp$) in the remainder of this article, we stick to the optimal choice of the angle $\phi=\frac{\pi}{2}$ ($\phi=\frac{\pi}{4}$) maximizing the signal.
For small angles $\vartheta'$ the signal photons induced by the ${\rm LG}_{1,0}$ high-intensity laser mode decay faster than those associated with the ${\rm LG}_{0
,0}$ mode. 
While the ${\rm LG}_{0,0}$ signal is monotonically decreasing to zero for larger $\vartheta'$, the ${\rm LG}_{1,0}$ signal exhibits an additional peak in the vicinity of $\vartheta'\approx70\,\mu{\rm rad}$.
The integrated total numbers of signal photons are $N^{\rm tot}\simeq23.77$ ($N^{\rm tot}\simeq11.57$) per shot for the case of the ${\rm LG}_{0,0}$ (${\rm LG}_{1,0}$) pump.
These numbers have to be compared with the huge number $N_{\rm in}\simeq10^{12}$ probe photons of the same energy, traversing the interaction region essentially unaltered and constituting the background from which the elastic signal is to be discriminated.
On the other hand, the number of signal photons contained within a ring around the beam axis delimited by $\vartheta'_{\rm min}=50\,\mu{\rm rad}$ and $\vartheta'_{\rm max}=100\,\mu{\rm rad}$ -- i.e., the angular interval where the side peak in the ${\rm LG}_{1,0}$ signal is located -- is $N^{\rm tot}_\circ\simeq1.31$ ($N^{\rm tot}_\circ\simeq4.50$), which is to be compared with the negligible number of probe photons $N_{{\rm in},\circ}\simeq1.14\times10^{-9}$ contained in the same angular interval.

For the perpendicularly polarized signal photons, the criterion for the principle possibility of their measurement with a polarimeter of polarization purity $\cal P$ is ${\rm d}N^\perp\geq{\cal P}{\rm d}N_{\rm in}$.
We call signal photons meeting this criterion {\it discernible}.
This criterion is fulfilled for signal photons scattered outside $\vartheta'_==22.36\,\mu{\rm rad}$ ($\vartheta'_==24.38\,\mu{\rm rad}$) for the ${\rm LG}_{0,0}$ (${\rm LG}_{1,0}$) pump.
The corresponding number of discernible signal photons per shot is $N^\perp_{\rm dis}\simeq0.61$ ($N^\perp_{\rm dis}\simeq0.34$).

Generically, we observe that the differential numbers of x-ray signal photons~\eqref{eq:dNtotperp_dcostheta} attainable in the head-on collision with a pure ${\rm LG}_{p,0}$ high-intensity  pump exhibits $p$ side peaks at finite values of $\vartheta'$.
Next-to-leading side peaks tend to be shifted to larger values of $\vartheta'$ and/or contain lower numbers of signal photons.
They might, however, be employed to induce interference phenomena in the all-optical parameter regime, reminiscent of those discussed for multi-beam configurations in Refs.~\cite{Hatsagortsyan:2011,King:2013am}.

Also note, that the fact that Eqs.~\eqref{eq:Ipp'}, \eqref{eq:dNtotperp_dcostheta} and \eqref{eq:dNin} depend on $\vartheta'$ only via the combination $w_0\omega\vartheta'$ implies that -- keeping all the other parameters fixed --- the results for the normalized angular decay of the signal and probe photons depicted in Fig.~\ref{fig:LG} can straightforwardly be rescaled to another x-ray photon energy $\omega^\ast$ as $\vartheta'\to\frac{\omega}{\omega^\ast}\vartheta'$.
The same is true for Figs.~\ref{fig:FGB} and \ref{fig:FGBGauss} discussed below.

\subsection{Flattened Gaussian beams}\label{subsec:FGB}

\begin{figure}[h]
 \centering
\includegraphics[width=0.67\columnwidth]{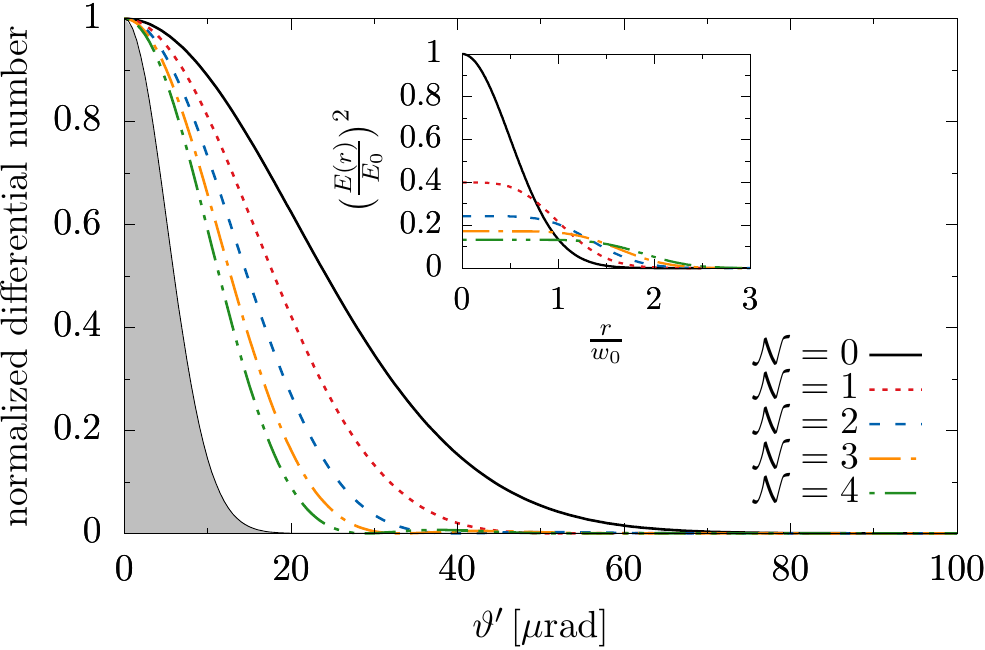}
\caption{Far-field angular decay of the signal~\eqref{eq:dNtotperp_dcostheta} and probe~\eqref{eq:dNin} (gray) for different transverse pump profiles.
For the high-intensity pump we consider different order $\cal N$ flattened Gaussian beams~\eqref{eq:EN} of the same pulse energy; see the inset for the radial intensity profiles in the focus.
The plot is for $\omega=12914\,{\rm eV}$, $w_{\rm probe}=3w_0$, $\lambda=800\,{\rm nm}$, $w_0=1000\,{\rm nm}$, $T=10\,{\rm fs}$ and $\tau=30\,{\rm fs}$.}
\label{fig:FGB}
\end{figure}

The class of flattened Gaussian beams (FGBs) encompasses infinitely many representatives, which are labeled by a positive integer $\cal N$ and defined via the focus profile \cite{Gori:1994}

\begin{equation}
 E_{\cal N}(r)=E_{0,{\cal N}}\,{\rm e}^{-(\frac{r}{w_0})^2}\sum_{k=0}^{\cal N}\frac{1}{k!}\Bigl(\frac{r}{w_0}\Bigr)^{2k} \,,
 \label{eq:EN}
\end{equation}
where $E_{0,{\cal N}}$ denotes the peak field amplitude.
Equation~\eqref{eq:EN} contains the fundamental Gaussian beam (${\cal N}=0$), and approaches a constant transverse field in the limit of ${\cal N}\to\infty$.
Generically speaking, the larger $\cal N$, the larger the effective waist $w_{\cal N}$ of the beam and the wider the flattened, plateau-like region about ${\rm z}=0$; cf.  Fig.~\ref{fig:FGB} for a graphical illustration.

An alternative representation of \Eqref{eq:EN} is \cite{Gori:1994}
\begin{equation}
 E_{\cal N}(r)=E_{0,{\cal N}}\,{\rm e}^{-(\frac{r}{w_0})^2}\sum_{p=0}^{\cal N} (-1)^p c_p L_p^0\Bigl(\bigl(\tfrac{\sqrt{2}r}{w_0}\bigr)^2\Bigr) ,
 \label{eq:E_G}
\end{equation}
with coefficients
\begin{equation}
 c_p:=\sum_{k=p}^{\cal N}\binom{k}{p}\frac{1}{2^k}>0\,.
\end{equation}

Equation~\eqref{eq:E_G} allows for the straightforward implementation of a flattened Gaussian high-intensity laser pulse of order $\cal N$ in the x-ray photon scattering scenario discussed above.
From the comparison of \Eqref{eq:E} specialized to $l=0$ and \Eqref{eq:E_G} we infer the following mapping
\begin{align}
 \mathfrak{E}_{0,p}\ \leftrightarrow\  E_{0,{\cal N}} c_p \quad\text{and}\quad
 \varphi_{0,p}\ \leftrightarrow\  p\pi\,, \label{eq:subs}
\end{align}
for $0\leq p\leq{\cal N}$ and $\mathfrak{E}_{0,p}=0$ for $p>{\cal N}$.
With the help of Eqs.~\eqref{eq:Equad} and \eqref{eq:W}, the peak field amplitude in \Eqref{eq:E_G} can then be expressed as
\begin{equation}
 E_{0,{\cal N}}^2\simeq 8\sqrt{\frac{2}{\pi}}\frac{W}{\pi w_0^2\tau}\frac{1}{C^2_{\cal N}}\,,\quad\text{with}\quad
 C^2_{\cal N}:=\sum_{p=0}^{\cal N}c_p^2=\sum_{k=0}^{\cal N}\sum_{l=0}^{\cal N}\binom{k+l}{k} \frac{1}{2^{k+l}}\,.
 \label{eq:E0quad}
\end{equation}
From \Eqref{eq:E0quad} we can infer that the effective waist of the flattened Gaussian beam of order $\cal N$ is proportional to $C_{\cal N}$, prompting us to define $w_{\cal N}=w_0C_{\cal N}$.
In turn, the energy put into the contributing Laguerre-Gaussian modes $(0,p)$ with $p\in\{0,\ldots{\cal N}\}$ is given by
\begin{equation}
 W_{0,p} \simeq \frac{c_p^2}{C^2_{\cal N}} W .
 \label{eq:W0p}
\end{equation}
Upon plugging Eqs.~\eqref{eq:subs} and \eqref{eq:W0p} into \Eqref{eq:Ipp'}, resulting in
\begin{align}
 \sum_{p,p'}{\cal I}_{p,p'}\to\ &\frac{W}{m_e}\sum_{p=0}^{\cal N}\sum_{p'=0}^{\cal N}\frac{c_pc_{p'}}{C^2_{\cal N}}\,(-1)^{p-p'}  \nonumber\\
  &\times\sum_{j=0}^p\sum_{j'=0}^{p'} \binom{p}{j}\binom{p'}{j'}\frac{\partial_c^{j+j'}}{ j!j'!} \frac{1}{c}F_{2|p-p'|}\Bigl(0,\tfrac{(w_0\omega\vartheta')^2}{8c}+\tfrac{(2{\rm z}_R)^2}{(\frac{T}{2})^2+\frac{1}{2}(\frac{\tau}{2})^2}\Bigr) \bigg|_{c=1},
  \label{eq:Ipp'FGB}
\end{align}
we can analyze the differential numbers~\eqref{eq:dNtotperp_dcostheta} of x-ray photons scattered at a focused high-intensity pulse of flattened Gaussian transverse profile.

In Fig.~\ref{fig:FGB} we focus on the far-field angular decay of the x-ray signal photons induced in the head-on collision of the x-ray probe with flattened high-intensity Gaussian pumps of different orders $\cal N$.
We limit ourselves to ${\cal N}\leq4$, implying that the considered flattened Gaussian focus profiles are -- to a good approximation -- uniformly illuminated by the x-ray probe of waist $w_{\rm probe}=3w_0$; see the inset in Fig.~\ref{fig:FGB}.
The general trend is as follows: the wider the pump, i.e., the larger $\cal N$, the narrower the scattering signal.

At the same time, both the integrated total numbers of signal photons $N^{\rm tot}$ and the numbers of discernible polarization-flipped signal photons $N^\perp_{\rm dis}$ diminish with $\cal N$:
\begin{center}
 \begin{tabular}{c||c|c|c|c|c|}
 ${\cal N}$ & $0$ & $1$ & $2$ & $3$ & $4$ \\
 \hline
 \hline
 $N^{\rm tot}$  & $\ 23.77\ $  & $\ 12.68\ $ & $\ 8.51\ $ & $\ 6.37\ $ & $\ 5.09\ $ \\
 \hline
 $N^\perp_{\rm dis}$ & $0.61$ & $0.17$ & $0.05$ &  $0.01$ & $0.005
$ \\
 \hline
\end{tabular}
\end{center}
This is a direct consequence of the fact that the peak field decreases with increasing $\cal N$.

\begin{figure}[h]
 \centering
\includegraphics[width=0.67\columnwidth]{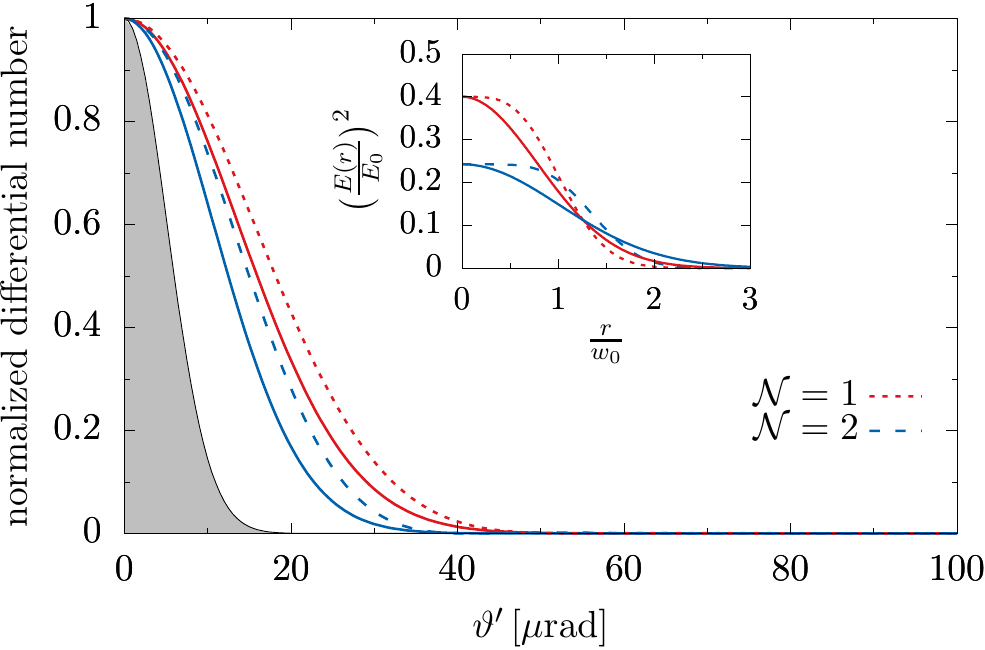}
\caption{Far-field angular decay of the signal~\eqref{eq:dNtotperp_dcostheta} and probe~\eqref{eq:dNin} (gray) for different transverse pump profiles.
Here, we compare the angular decay of the signal for flattened Gaussian pumps~\eqref{eq:EN} of order ${\cal N}\in\{1,2\}$ (dashed lines) with the analogous results for a fundamental Gaussian pump with rescaled waist $w_0\to w_{\cal N}$ (solid lines); see the inset for the respective focus profiles. The plot is for $\omega=12914\,{\rm eV}$, $w_{\rm probe}=3w_0$, $\lambda=800\,{\rm nm}$, $w_0=1000\,{\rm nm}$, $T=10\,{\rm fs}$ and $\tau=30\,{\rm fs}$.}
\label{fig:FGBGauss}
\end{figure}

Finally, in Fig.~\ref{fig:FGBGauss} we compare our FGB results for the representative orders ${\cal N}\in\{1,2\}$ with the analogous results obtained for Gaussian pump fields of rescaled width $w_0\to w_{\cal N}$.
This is of special interest as it allows for an assessment of the role of the details of the transverse focus profile of the pump on the angular decay of the signal photons.
Figure~\ref{fig:FGBGauss} shows that the angular decay of the scattering signal associated with an FGB profile of order $\cal N$ is slightly less pronounced than that for the corresponding rescaled Gaussian pump. 
This indicates that the steepness of the transverse focus profile of the pump impacts the angular decay of the signal: a steepening of the pump profile while keeping its peak field and pulse energy fixed seems to result in a less pronounced angular fall-off. 
The signal photon numbers associated with the rescaled Gaussian pump highlighted in Fig.~\ref{fig:FGBGauss} are $N^{\rm tot}\simeq10.23$ ($N^{\rm tot}\simeq6.26$), $N_{{\rm dis}}^\perp\simeq0.11$ ($N_{{\rm dis}}^\perp\simeq0.02$) for ${\cal N}=1$ (${\cal N}=2$). These values are of the same order, but somewhat smaller than the corresponding FGB results given above.
Correspondingly, our results indicate that for pumps featuring transverse focus profiles of similar characteristics (same pulse energy and peak field, similar waist spot size, monotonic fall-off with $r$) the scattering phenomenon of signal photons outside the forward cone of the probe is not very sensitive to the details of the pump's transverse focus profile.
This is reassuring for the perspectives of searching this effect in experiment: as to date the precise intensity distribution of focused high-intensity laser pulses amount to critical unknowns, the stability of the scattering phenomenon under deformations of the pulse profile is of central importance for the feasibility of such an experiment with state-of-the-art technology.
On the other hand, our present analysis shows that a quantitative precision study of QED vacuum fluctuation mediated x-ray photon scattering at a focused high-intensity laser pulse invitably requires knowledge about the particular focus profile available in experiment.

\section{Conclusions and Outlook}\label{sec:concls}

In this article we generalized Ref.~\cite{Karbstein:2015xra} studying vacuum birefringence in the head-on collision of a weakly focused x-ray probe with a tightly focused high-intensity pump, described as fundamental paraxial Gaussian mode, to paraxial Laguerre-Gaussian beams of arbitrary mode composition.
In this context, we explicitly showed that the total number of signal photons attainable in a polarization insensitive measurement is maximized for a different choice of the polarization vectors of the colliding beams than the number of polarization flipped signal photons constituting the vacuum birefringence signal.
Resorting to a few well-justified approximations, we were able to provide simple analytical estimates for the differential numbers of signal photons as a function of the polar angle $\vartheta'$ measured from the forward beam axis of the x-ray probe.
Finally, we discussed some exemplary results and emphasized the differences from the sceanario with a fundamental Gaussian pump.
More specifically, we focused on pump fields corresponding to a pure Laguerre-Gaussian mode and to the class of flattened Gaussian beams~\cite{Gori:1994}.
From the comparison of the far-field angular decay of the signal photons induced by a flattened Gaussian pump and a fundamental Gaussian pump of similar waist size in the focus we inferred that -- keeping the pulse energy and  the peak field the
same -- the scattering signal is not very sensitive to the details of the pump's transverse focusing profile.

The formulas presented in the present article allow for the study of vacuum-fluctuation-mediated x-ray photon scattering at head-on colliding, rotationally symmetric paraxial high-intensity laser pulses of arbitrary transverse focus profiles.
By construction they are limited to the parameter regime where $w_{\rm probe}\gtrsim w_0$.
Their generalization to finite collision angles as well as elliptical focus profiles of the pump is straightforward.
We are confident that our results will stimulated further investigations of the fate of the x-ray scattering signal under deformations of the driving laser pulses in the interaction volume.
Interesting directions would be the study of similar effects in high-intensity laser pulse collisions in the all-optical regime, and explicitly accounting for the transverse profile of the x-ray probe.

\acknowledgments

This work has been funded by the Deutsche Forschungsgemeinschaft (DFG) under Grant No. 416607684 within the Research Unit FOR2783/1 (FK) and is supported by Russian Science Foundation grant RSCF 17-72-20013 (EM).

\end{document}